\documentclass[12pt]{spieman}  
\usepackage{amsmath,amsfonts,amssymb}
\usepackage{graphicx}
\usepackage{setspace}
\usepackage{tocloft}
\usepackage{array}
\usepackage{booktabs}
\usepackage{float}
\usepackage{multirow} 
\title{Impact-Induced Cartilage Damage Assessed using Polarisation-Sensitive Optical Coherence Tomography}

\author[a,b,*]{Matthew Goodwin}
\author[c]{Joshua Workman}
\author[c]{Ashvin Thambyah}
\author[a,b]{Fr\'{e}d\'{e}rique Vanholsbeeck}

\affil[a]{The Dodd-Walls Centre for Photonic and Quantum Technologies, The University of Auckland, Auckland, New Zealand 1010.}
\affil[b]{Department of Physics, The University of Auckland, Auckland, New Zealand, 1010.}
\affil[c]{Department of Chemical and Materials Engineering, The University of Auckland, Auckland, New Zealand, 1010.}

\cftpagenumbersoff{figure}
\cftpagenumbersoff{table} 
\begin{document} 
\maketitle

\begin{abstract}

\noindent
\textbf{Significance}: Non-invasive determination of structural changes in articular cartilage immediately after impact and rehydration provides insight into the response and recovery of the soft tissue, as well as provides a potential methodology for clinicians to quantify early degenerative changes.

\noindent
\textbf{Aim}: To apply Polarisation-Sensitive Optical Coherence Tomography (PS-OCT) to examine subtle alterations of the optical properties in healthy and early-stage degenerate articular cartilage immediately after impact loading to identify structurally relevant metrics required for understanding the mechanical factors of osteoarthritic initiation and progression.

\noindent
\textbf{Approach}: A custom-designed impact testing rig was used to deliver 0.9 J and 1.4 J impact energies to bovine articular cartilage. A total of 55 (n=28 healthy, n=27 mildly degenerate) cartilage-on-bone samples were imaged before, immediately after, and 3 hours after impact. PS-OCT images were analysed to assess changes relating to surface irregularity, optical attenuation, and birefringence.

\noindent
\textbf{Results}: Mildly degenerate cartilage exhibits a significant change in birefringence following 1.4 J impact energies compared to healthy samples which is believed to be attributable to degenerate cartilage being unable to fully utilise the fluid phase to distribute and dampen the energy. After rehydration, the polarisation-sensitive images appear to `optically-recover' reducing the reliability of birefringence as an absolute metric. Surface irregularity and optical attenuation encode diagnostically relevant information and may serve as markers to predict the mechanical response of articular cartilage.

\noindent
\textbf{Conclusion}: Subtle alterations in the collagen network associated with early stage degeneration can influence the mechanical response to impact loading. PS-OCT with its ability to  non-invasively image the sub-surface microstructural abnormalities of cartilage presents as an ideal modality for cartilage degeneration assessment and identification of mechanically vulnerable tissue.
\end{abstract}


{\noindent \footnotesize\textbf{*}Matthew Goodwin,  \linkable{Matthew.Goodwin@auckland.ac.nz} }

\begin{spacing}{2}   

\section{Introduction}
\label{sect:intro}  

The interaction between the ultra-structure of the collagen network, proteoglycan complexes and interstitial fluid result in articular cartilage (AC) being capable of resisting a range of dynamic loads and contribute to its poro-viscoeleastic properties. Fluid movement during high strain rates is restricted, enabling the tissue to withstand impact loading. Previous studies have shown that, during early stages of degeneration, the disruption of the extracellular matrix leads to an increase in water content~\cite{guilak1994mechanical} and alters the matrix permeability~\cite{makela2015very}. The culmination of these aspects result in reduced mechanical functionality and are believed to be key mechanical factors for the initiation and progression of osteoarthritis (OA)~\cite{saarakkala2010depth}.

OA is considered commonly as a disease of age, where the loss of articular cartilage is gradual and linked to  factors such as genetics, obesity, occupational risk and diet~\cite{vina2018epidemiology,zhang2010epidemiology}. Due to the aging population of Western society, the sociological and economic burden of OA has necessitated research into understanding the pathophysiology and development of new diagnostic and treatment methods. Unfortunately, clinical diagnosis of OA occurs well into the late stages of the disease, where the degeneration is irreversible and the treatment options are limited to analgesics, joint replacement, and joint fusion~\cite{buckwalter2008operative, moskowitz2007osteoarthritis}. Post-traumatic OA (PTOA) is more common in young-to-middle-aged individuals and often is a result of acute joint trauma where excessive forces experienced in the joint can lead  to chondral matrix damage, cartilage disruption, and/or subchondral bone fracture~\cite{buckwalter2004joint}. The limited lifetime of the aforementioned joint replacement and the reduction in quality of life linked with joint fusion makes these treatment options less ideal for young and middle-aged adults. Therefore, detecting the early stages (pre-OA) of degeneration is crucial and while several novel therapy methods exist~\cite{chikanza2000novel,evans2005novel,krishnan2005novel}, detecting this narrow window in the disease spectrum remains increasingly elusive.

In the research and pre-clinical domain, arthroscopy is considered the gold standard for early degenerative lesion assessment~\cite{chu2010clinical, chu2012early, kijowski2006radiographic, santori1999arthroscopic}, however only surface features are detectable and its effectiveness remains questionable~\cite{spahn2009valid, spahn2011reliability}. A standard clinical diagnosis is made by examining a weight-bearing X-ray of the patient, where narrowing of the joint space is an indicator of cartilage erosion and OA~\cite{laurin1979tangential,hart1991clinical,balint1996diagnosis}. Optical coherence tomography (OCT) has emerged as a potential imaging modality due to its non-invasive, non-radiative, high-resolution and real-time nature. OCT  bridges the gap between non-invasive, macroscopic resolution imaging modalities used in clinical settings and destructive, high resolution research-centric techniques such as microscopy. 

OCT entered the OA research domain in the late 90s where Herrmann et al.~\cite{herrmann1999high} demonstrated the ability of OCT to detect microstructural changes associated with OA and hinted at the diagnostic value the modality may hold. In the early 2000s~\textit{in vitro} experiments proved the potential of OCT, launching several arthroscopic based \textit{in vivo} trials with promising results~\cite{pan2003hand,han2003analysis,chu2004arthroscopic,li2005high,patel2005monitoring}. Adding polarisation-sensitivity to OCT (PS-OCT) allowed another degree of contrast and enabled the extraction of information regarding the collagen network~\cite{drexler2001correlation,ugryumova2009novel}. Furthermore, PS-OCT aids in the identification of sub-surface degenerative changes related to the extracellular matrix, increasing the diagnostic value of the technique~\cite{li2005high,xie2006determination,chu2007clinical}. Over the last decade the focus has been on developing quantitative metrics to objectively grade and characterize degeneration. Currently, there are intensity-based metrics which quantify surface reflectively~\cite{saarakkala2009quantification}, surface integrity, matrix homogeneity, and optical attenuation~\cite{marx2012vitro,nebelung2014morphometric,brill2016polarization}. PS-OCT based metrics revolve around quantifying the strength of the birefringence - an indirect measure of collagen matrix arrangement~\cite{brill2016polarization,goodwin2018quantifying}. Optical axis orientation can also be extracted from polarisation-sensitive measurements which has recently allowed the full three dimensional reconstruction of collagen fiber orientation using a tractographic extension of PS-OCT~\cite{yao2016nondestructive,wang2016mapping}. Currently, de Bont et al.~\cite{de2015evaluation} are the only group to have parameterised impact-induced cartilage degeneration using OCT and while they concluded OCT can reliably detect degenerative features, no polarisation metrics have been evaluated nor has the optical response of degenerate cartilage been evaluated following impact. It is also important to note that the previous OCT study was carried out only on tissue obtained from OA patients, which leaves open the question of how healthy and structurally intact cartilage may respond. Bovine cartilage tissue from young adult bulls provides a source of pristine quality cartilage that acts as a suitable model to study healthy cartilage mechanics. Further, with the early-OA model of bovine cartilage degeneration, validated against human OA tissue~\cite{hargrave2013bovineModel}, a systematic analysis of the potential use of PS-OCT to study how microstructural degeneration influences cartilage tissue mechanics can be carried out.

Therefore, the objective of this study  is to explore the response and recovery of healthy and degenerate bovine cartilage following impact loading  using PS-OCT.

\section{Materials and Methods}
\subsection{Classification and Preparation of Cartilage Samples.}
\begin{figure}[htbp]
  \centering
  \includegraphics[width=0.8\linewidth]{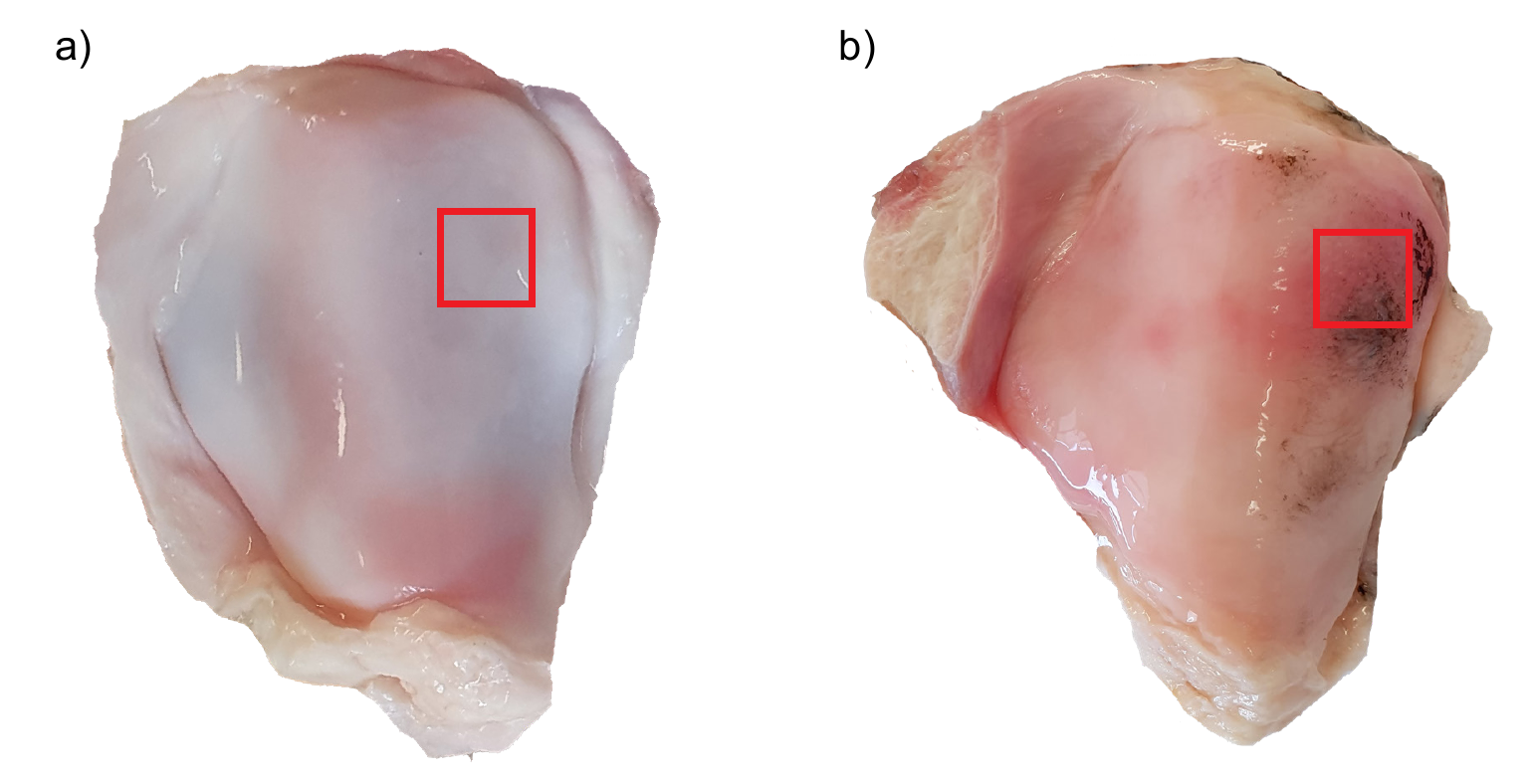}
\caption{ Healthy (a) and mildly degenerate (b) Bovine patella with the distal lateral region indicated (red rectangle). India ink completely washes off healthy articular cartilage, indicating an intact surface layer whereas fibrillation as a result of degeneration will retain the ink. }
\label{fig:patella}
\end{figure}

Bovine patella (n=55) were chosen for this study due to their near limitless supply and recent validation as a model for studying early OA~\cite{hargrave2013bovineModel}. The patella were sourced from a local slaughterhouse and stored at -20$^\circ$C until use. Each patella only underwent a single freeze-thaw cycle that has been previously shown to have no significant effects on the biomechanical properties of the tissue~\cite{changoor2010effects,kiefer1989effect,szarko2010freeze}.  Samples were thawed under running water for at least 30 minutes before 12 mm$^{3}$ cartilage-on-bone sections were extracted from the distal-lateral region. A mark was etched into the distal region of the bone to ensure impact, OCT, and analysis were all performed with the same orientation.  The distal-lateral area has been identified as a region where degeneration is first observed and also provides a flat sample area which is ideal for impact testing~\cite{thambyah2010subtle}.
The samples were macroscopically graded using India ink according to the Outerbridge classification system~\cite{outerbridge1961etiology}. Mildly degenerate tissue was obtained from patellae containing surface fibrillation, indicated by a mild positive staining with India ink~\cite{meachim1972light}. Pristine healthy tissue showed no sign of surface disruption and India ink washed off completely.

Each sample was placed in a custom stainless steel sample holder and fixed in place using dental plaster (Moldastone, Kulzer, Hanau, Germany), ensuring that the cartilage surface was parallel to impact tip. During the plaster setting time ($\approx$12 minutes) the samples were kept hydrated with 0.15M saline and, once set, were immersed in saline and only removed for a brief period ($<$2 minutes) for each impact and imaging events.

\subsection{Impact Loading.}
\begin{figure}[htbp]
  \centering
  \includegraphics[width=0.3\linewidth]{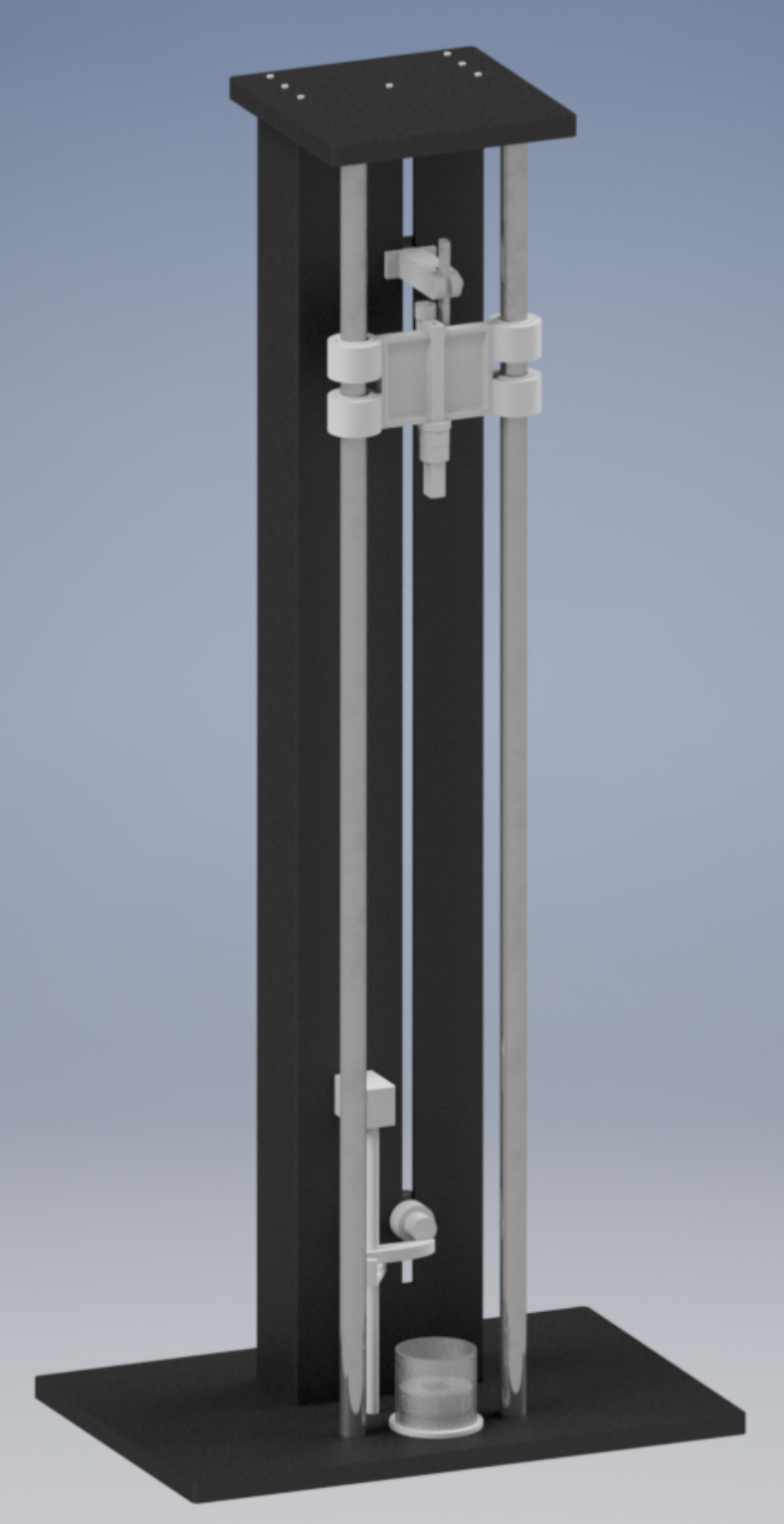}
\caption{Schematic of the Impact tower.}
\label{fig:impact}
\end{figure}

A single impact event was carried out using a custom-built impact rig (figure~\ref{fig:impact}) which has been described previously~\cite{workman2017influence}. Briefly, the rig consists of a vertical moving impact assembly which slides along two stainless steel rails using Teflon bushings. The assembly is equipped with a 10~mm diameter stainless steel impact tip which can be released from various heights to adjust the impact energies. A 9021A piezoelectric force transducer (Kistler Instrument Ltd., AG, Schwuiz, Germany) was attached to the assembly, allowing the peak stress to be extracted. The transducer output was passed into a charge amplifier (Model 5015A, Kistler AG.) and sampled at 1 MS/s using a USB DAQ (USB-1208HS-2AO, Measurement Computing Corporation, MA, USA).

The impact was video recorded at 2,000 fps using a high speed camera (FASTCAM MC2, Photron, Tokyo, Japan) equipped with a 75mm F1.4 lens (Computar, CBC Americas Corp., Cary, NC, USA)  and a low heat LED. The output video underwent automated frame-by-frame motion analysis, allowing impact (\textit{$v_{impact}$}) and rebound velocity (\textit{$v_{rebound}$}) to be determined using Photron FASTCAM analysis software. The kinetic energy lost during the impact enabled the coefficient of restitution (\textit{CoR}) to be calculated:

\begin{equation}
    CoR = \frac{v_{rebound}}{v_{impact}}
\end{equation}
The cartilage samples were split into two impact groups: energy level 1 (EL1) and energy level 2 (EL2) with impact energies of 0.9 J and 1.4 J respectively. A summary of the impact parameters can be found in table~\ref{table:impact values}. 

\begin{table}[htbp]
\centering
\caption{Summary table of the impact parameters measured during the single impact event.}
\begin{tabular}{  c | c| c  } 
\hline
 Drop Height (mm) & Impact Energy (J)  & Impact Group \\
\hline
160 & 0.9  & EL1\\ 
\hline
250 & 1.4  & EL2\\ 
\hline
\end{tabular}
\label{table:impact values}
\end{table}

\subsection{OCT and Image Analysis}
A custom PS-OCT system, described previously~\cite{thampi2020towards}, was used for the imaging. Briefly, a swept-source system with a center wavelength of 1310~nm, full width half maximum bandwidth of 100~nm, and 50~kHz repetition rate  was used. All samples were imaged at three different time points: before the impact (pre), immediately after the impact  event (post), and 3 hours after the event (rehydration). Volumetric C-scans were taken such that the impact region and the surrounding non-impacted were recorded.  Quantitative analysis was performed using a combination of LabVIEW and python scripts to assess the optical parameters relating to cartilage degeneration. For each instance, a volumetric image with five times averaging underwent segmentation to isolate the impact area for analysis.

Three optical parameters were calculated: surface irregularity, attenuation index and the cumulative birefringence coefficient. The former two are derived from the traditional intensity-based OCT data and the latter utilises the polarisation-sensitivity of the system. Surface irregularity is a measure of the roughness of the cartilage surface and has been used in a number of studies previously~\cite{nebelung2014morphometric,nebelung2015three}. Briefly, it is calculated by first performing surface detection through thresholding and determining the maximum value of each A-scan. The detected surface is then compared to a smoothed surface through the use of a Savitzky-Golay filter and the irregularity is calculated as the standard deviation between the actual and fitted values on a B-scan basis. A higher surface irregularity is an indicator of surface fibrillation and/or discontinuity. Attenuation index relates to the absorption and scattering properties of a material and has been utilised in several studies~\cite{brill2016polarization,de2015evaluation,nebelung2015three}. In the most simple form, it can be calculated by creating a linear fit for each A-scan intensity profile, where a larger value reflects an increase in optical attenuation. The cumulative birefringence coefficient measures the strength of the birefringence by calculating the retardation gradient of each A-scan~\cite{brill2016polarization,goodwin2018quantifying}. A small spatial averaging window and Gaussian filter is applied to each A-scan to smooth the retardation data before the cumulative value was calculated. Larger gradient values indicate the tissue exhibits stronger birefringence. 

\subsection{Statistical Analysis}
The Kruskal-Wallis non-parametric test followed by Dunn's post hoc test with Holm-Bonferroni corrections was used to identify any significant differences between health states, impact energies and time points. \textit{p-values} of $p\leq 0.05$ were considered statistically significant and highlighted in the results section depending on the level of significance ([ns] = non-significant $[{*}]: 0.01 \leq p \leq 0.05; [{**}]: 0.001 \leq p < 0.01; [{***}]: p < 0.001$). Statistical testing was carried out in Python using the \textit{scipy} and \textit{scikit-posthocs} modules.

\subsubsection{Data visualisation} Principal component analysis (PCA) is a dimensionality reduction technique that allows high-dimensional datasets to be reduced to a few 'principal components` which contain the important information~\cite{abdi2010principal}. PCA was employed to evaluate the classification ability and highlight the effectiveness of each quantitative metric to differentiate healthy and degenerate tissue in the pre-impact state. For every volumetric image,  PCA was performed on the histograms of the quantitative OCT metric using the \textit{scikit-learn} module in Python~\cite{scikit-learn}.

\section{Results}

\subsection{Impact peak force and velocity}
\begin{figure}
\centering
  \includegraphics[width=\linewidth]{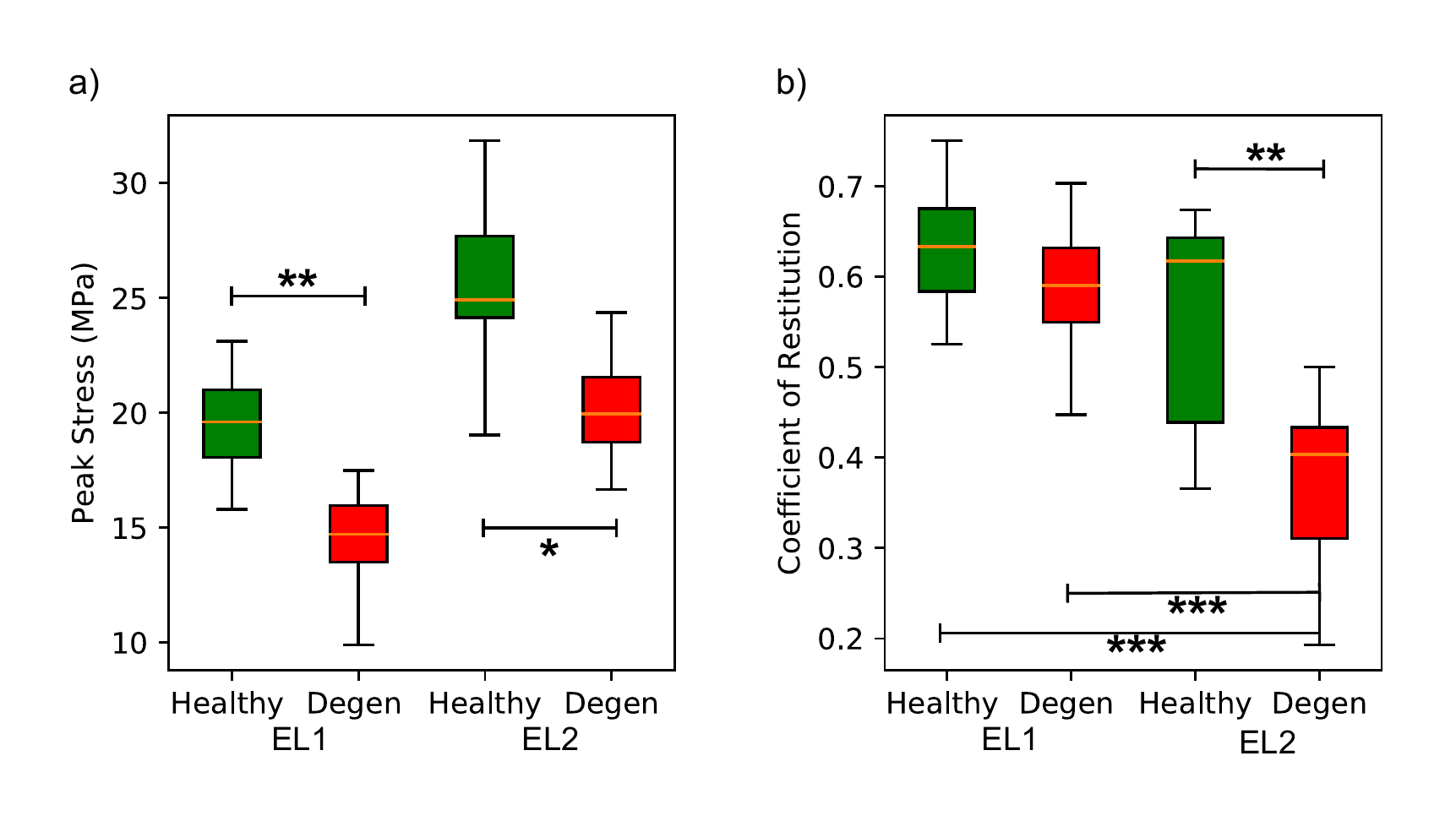}
  \caption{Peak impact stress data illustrating the significant decrease in peak force associated with degenerate samples for both EL1 and EL2 impact energies (a). Coefficient of restitution is significant reduced for degenerate samples experiencing EL2  impact events while all other samples have no significant changes regardless of health state or impact energy (b).}
  \label{fig:ForceVelocityImage}
\end{figure}

Samples with cartilage showing signs of early degeneration responded with a significant reduction in peak stress compared to healthy samples for the same impact energy (Figure~\ref{fig:ForceVelocityImage}a). While there is a monotonic decrease in the mean coefficient of restitution with cartilage degeneration, these differences were only significant for the higher impact energies, indicating a larger fraction of kinetic energy was deposited in the tissue~(Figure~\ref{fig:ForceVelocityImage}b). A summary of the results can be found in table~\ref{tab:ResultData}. 

\begin{table}
    \begin{center}
    \caption{Force, velocity and quantitative OCT results for each impact energy and cartilage health state. Values are expressed as mean $\pm$  standard deviation.}
    \renewcommand{\arraystretch}{1.8}
    \begin{tabular}{p{3cm} c|c|c|c|c}
    \hline
    \hline
     & & \multicolumn{4}{c}{Impact Energy}\\
     & & \multicolumn{2}{c}{Energy Level 1}&\multicolumn{2}{c}{Energy Level 2}\\
    \hline
     & & Healthy  & Degen & Healthy & Degen\\
     & & (n=13)  & (n=13) & (n=15) & (n=14)\\
    \hline
    \hline
    \multicolumn{2}{c|}{Peak Stress (MPa)} & 19.3~$\pm$~2.4 & 14.5~$\pm$~1.8 & 25.3~$\pm$~3.9 & 20.4~$\pm$~2.4\\
    \multicolumn{2}{c|}{Coefficient of Restitution (CoR)} & 0.63~$\pm$~0.06 & 0.57~$\pm$~0.07 & 0.55~$\pm$~0.11 & 0.38~$\pm$~0.08\\
    \hline
    \hline
    \multirow{3}{*}{\parbox{3cm}{Attenuation index (A.U.)}} & Pre-Impact & 0.31~$\pm$~0.06 & 0.54~$\pm$~0.06 & 0.30~$\pm$~0.12 & 0.51~$\pm$~0.08\\
     & Post-Impact & 0.36~$\pm$~0.04 & 0.50~$\pm$~0.11 & 0.32~$\pm$~0.09 & 0.46~$\pm$~0.09\\
     & Rehydration & 0.39~$\pm$~0.07 & 0.53~$\pm$~0.17 & 0.39~$\pm$~0.10 & 0.46~$\pm$~0.11\\
    \hline
    \multirow{3}{*}{\parbox{3cm}{Cumulative  birefringence  ($c\Delta n)\times~10^{-4}$}} & Pre-Impact & 0.93~$\pm$~0.20 & 1.04~$\pm$~0.13 & 0.92~$\pm$~0.14 & 1.05~$\pm$~0.13\\
     & Post-Impact & 0.91~$\pm$~0.23 & 1.13~$\pm$~0.23 & 0.99~$\pm$~0.13 & 1.63~$\pm$~0.32\\
     & Rehydration & 0.89~$\pm$~0.19 & 1.04~$\pm$~0.28 & 0.89~$\pm$~0.11 & 1.20~$\pm$~0.26\\
    \hline
    \multirow{3}{*}{\parbox{3cm}{Surface Irregularity (A.U.)}} & Pre-Impact & 1.44~$\pm$~0.32 & 4.01~$\pm$~0.77 & 1.22~$\pm$~0.27 & 4.65~$\pm$~1.73\\
     & Post-Impact & 4.21~$\pm$~1.28 & 5.69~$\pm$~2.64 & 7.96~$\pm$~2.84 & 7.71~$\pm$~3.78\\
     & Rehydration & 3.30~$\pm$~0.85 & 5.65~$\pm$~1.81 & 7.79~$\pm$~4.13 & 10.70~$\pm$~5.56\\
    \hline
    \hline
    \end{tabular}
    \label{tab:ResultData}
    \end{center}
\end{table}

\subsection{OCT Results}

\begin{figure}
\centering
  \includegraphics[width=\linewidth]{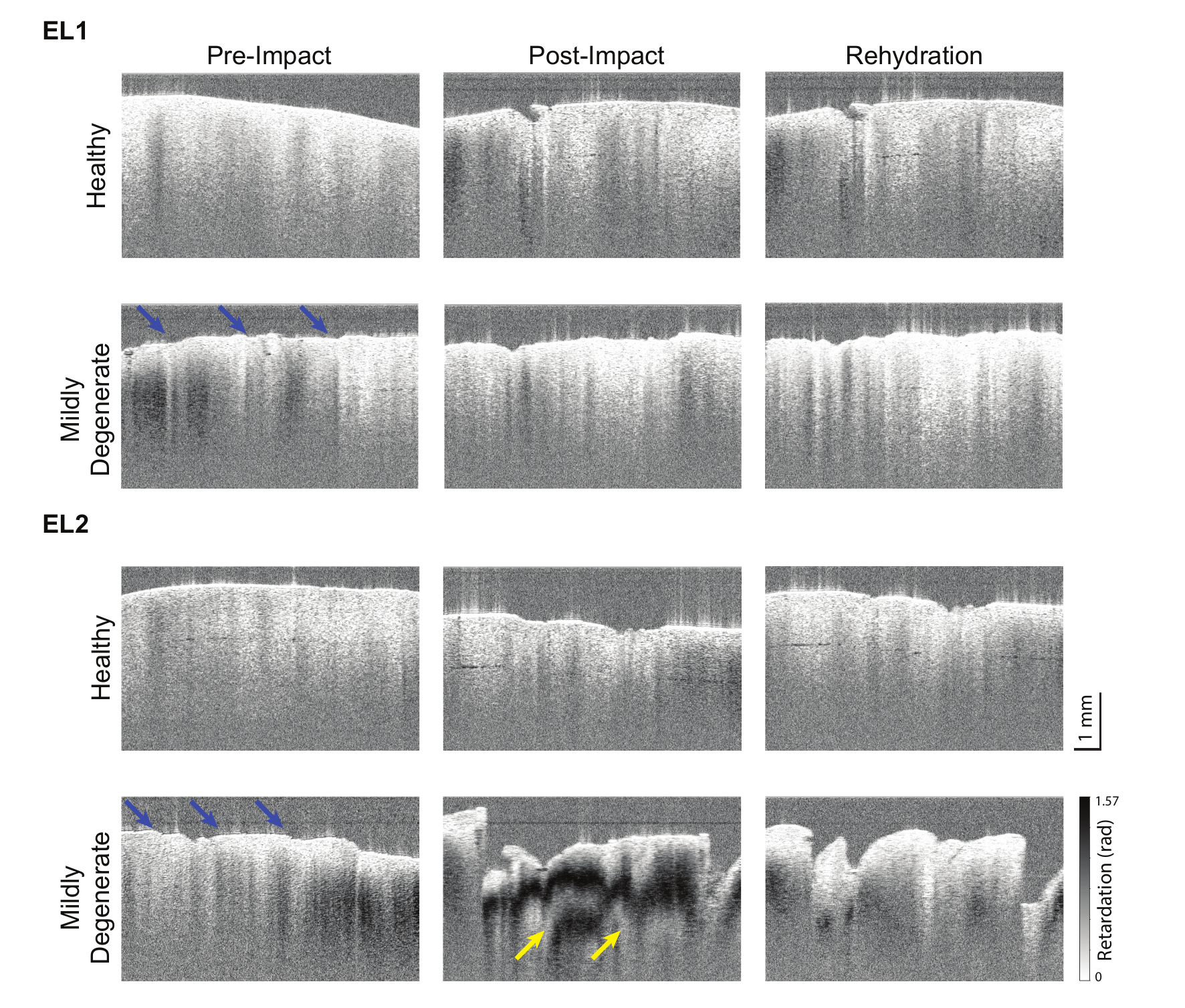}
  \caption{Polarisation-sensitive OCT images for healthy and mildly degenerate samples taken pre-impact, post-impact  and 3 hours after impact (rehydration) for both EL1 and EL2 impact events. Blue arrows highlight the natural surface fibrillation present in the degenerate samples. Yellow arrows highlight the area of increased optical activity.}
  \label{fig:PSOCT}
\end{figure}

\begin{figure}[htb]
  \includegraphics[width=\linewidth]{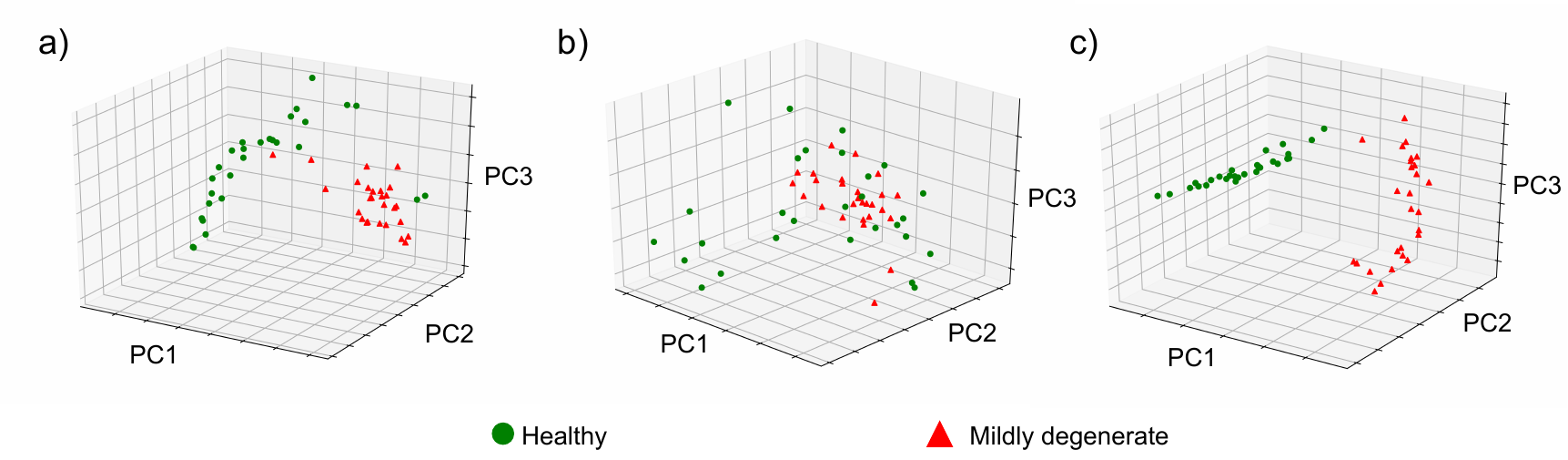}
  \caption{PCA results for attenuation (a), birefringence (b) and surface irregularity (c) for all pre-impact healthy (green) and degenerate (red) samples. We see clear clustering and separation of the two health states when analysed using the attenuation index and surface irregularity. However, birefringence shows little grouping in the first 3 principal components.}
  \label{fig:PCAPre}
\end{figure}

The intensity images reveal a homogeneous scattering matrix with the key difference between pre-impacted samples encoded in the congruity of the surface; healthy samples have a smooth, intact superficial matrix layer while surface fibrillation is present in all degenerate samples which is consistent with the macroscopic grading technique (Figure~\ref{fig:PSOCT}). Pre-impacted healthy samples were found to have a significant reduction [{*}] in the surface irregularity compared to the corresponding degenerate samples. The results from the PCA reflect this finding as a clear separation plane exists between healthy and degenerate samples (Figure~\ref{fig:PCAPre}c). Following EL1 impact, surface irregularity did increase for both healthy and degenerate samples, however, the increase was found to be non-significant [ns] even after rehydration. EL2 impact events resulted in a significant increase in surface irregularity, but only in healthy samples [{***}]. After rehydration of the matrix, the surface irregularity of healthy samples subjected to EL2 impact events remained significantly different from their original pre-impact state [{***}].

Pre-impacted healthy cartilage exhibited a significantly lower [{*}] attenuation index compared to degenerate cartilage. Figure~\ref{fig:PCAPre} (a) highlights the difference in attenuation index, with degenerate samples clustering together reasonably well. Immediately after impact healthy cartilage still had a significantly lower [{*}] attenuation index regardless of the impact energy. Interestingly, rehydrated healthy samples experienced an increase in the attenuation index and were no longer significantly different [ns] to degenerate cartilage. There were no significant [ns] changes in the attenuation index within health states between the measured time points (pre, post, rehydration).

Quantitative analysis of the polarisation-sensitive images revealed there is no significant difference [ns] in cumulative birefringence between pre-impacted healthy and mildly degenerate cartilage (Figures~\ref{fig:PSOCT}). In addition, both cartilage groups withstood EL1 impacts, showing no change [ns] in birefringence immediately after impact and following rehydration of the matrix. Healthy samples exhibited no change [ns] in birefringence following EL2 impact events or after rehydration. However, post-impact mildly degenerate samples suffered matrix failure leading to a significant increase in birefringence compared to pre-impact healthy [{***}] and mildly degenerate samples [{*}].  What is of particular interest is that degenerate samples `optically recovered' (to and extent) during rehydration. This `recovery' leads to the rehydrated samples being no longer significantly different [ns] from pre- or post-impact mildly degenerate samples via birefringence analysis. The reduction in birefringence during this period leads to these rehydrated samples being not significantly different [ns] to pre-impact or post-impact degenerate samples. 

\begin{figure}
\centering
  \includegraphics[width=\linewidth]{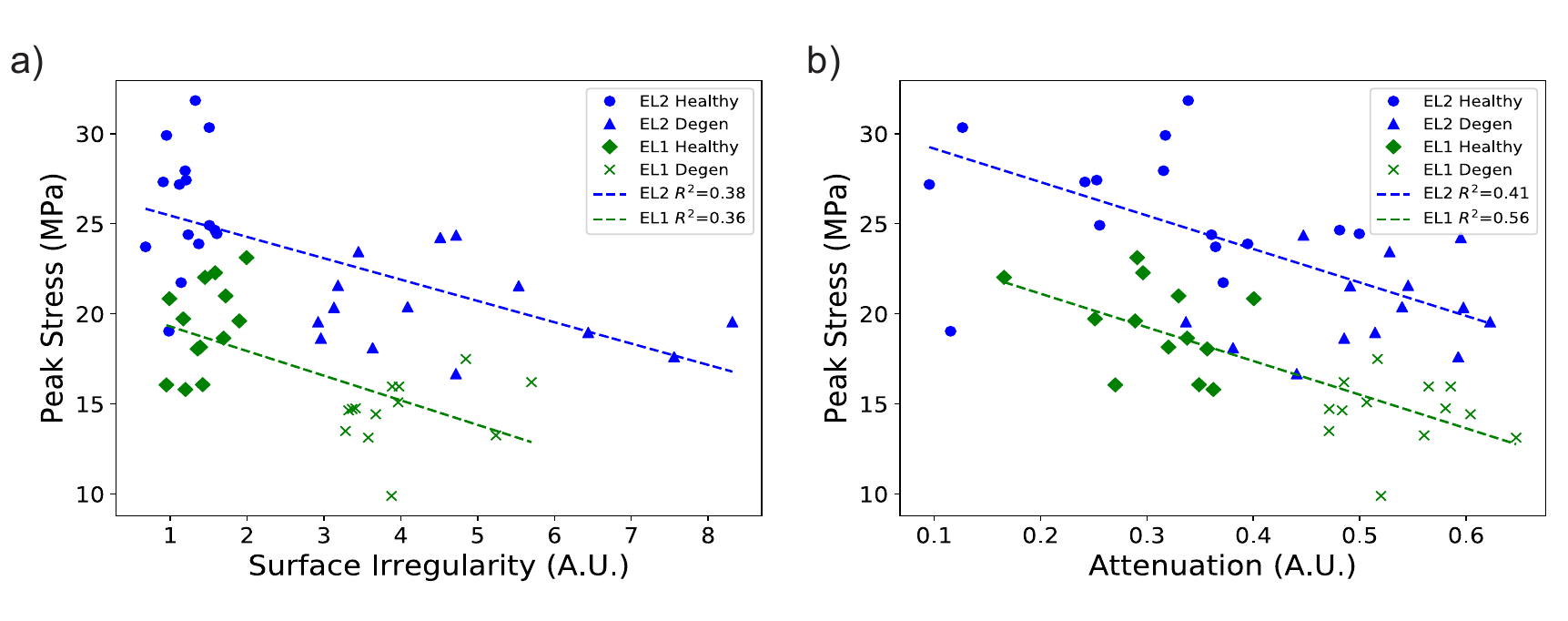}
  \caption{A weak correlation exists between the peak impact stress and pre-impact surface irregularity for both impact energies (a). Simiarly, a weak-moderate correlation is present between peak impact stress and pre-impact attenuation for EL1 and EL2 impact events (b).}
  \label{fig:ForceOCT}
\end{figure}

For both EL1 and EL2 impact events, weak correlations were identified between the peak impact stress and two of the pre-impact metrics: surface irregularity (EL1 $R^{2} =0.36$, EL2 $R^{2} =0.38$)  and attenuation (EL1 $R^{2} =0.56$, EL2 $R^{2} =0.41$) as shown in figure~\ref{fig:ForceOCT}.

\section{Discussion}
The most important finding of the study is the ability of PS-OCT to capture the dynamic and static changes occurring in healthy and degenerate bovine cartilage during impact and rehydration. Quantitative analysis using conventional, intensity-based OCT metrics generally agreed with previous studies which reinforces the potential of conventional OCT to be used as a diagnostic tool. Surface irregularity still remains one of the more reliable quantification metrics for assessing early degeneration as demonstrated with the clear classification plane present in Figure~\ref{fig:PCAPre}c.

Intuitively, surface irregularity increases with degeneration and impact energy. The formation of cracks has been proposed as a method for dissipating impact energy~\cite{fulcher2009viscoelastic} and can be seen in the post impact images in figure~\ref{fig:PSOCT}. The lack of congruity in the surface layer has been linked to accelerated degeneration and while quantifying this is useful in itself, identifying biomarkers that appear earlier in the degeneration process may be of greater diagnostic value.~\cite{pritzker2006osteoarthritis}

One of the earliest signs of degeneration during disease onset is a softening and swelling of the tissue.\cite{setton1994mechanical} Therefore, the question raised is: does OCT have the required sensitivity to detect changes in attenuation that correspond to an increase in water content? While Brill et al.~\cite{brill2016polarization} noted that degenerate cartilage exhibits higher attenuation, the differences only become significant during later stages of the disease. Our results indicate that there is a weakly significant increase in attenuation in mildly degenerate cartilage. Further, the negative correlation that exists between the attenuation value and the peak impact stress supports the notion of softening and swelling and highlights the potential of OCT as a method to non-invasively predict the mechanical response of cartilage to impact.
In terms of time-based differences, our results aligned with the study from de Bont et al.\cite{de2015evaluation}; we found no significant increase in attenuation directly following impact or with rehydration, indicating conventional OCT may not have the sensitivity to detect water exudation- at least not using 1300 nm.

Over the last two decades, several studies have attempted to explore the benefit of acquiring polarisation-sensitive data to provide an additional contrast for assessing cartilage degeneration. There have been conflicting findings with some groups associating the presence of birefringence with cartilage degeneration~\cite{xie2006determination} whereas others suggest natural, healthy cartilage should appear anisotropic~\cite{li2005high, chu2007clinical}. The contradictory findings have generally been reconciled by examination of the collagen fiber orientation and topographical variations that exists in joint~\cite{ugryumova2009novel,xie2006use,xie2008topographical}. In agreement with the other PS-OCT studies using bovine samples, healthy and degenerate cartilage exhibit little-to-no birefringence, and are not significant from one another~\cite{goodwin2018quantifying,xie2006determination,ugryumova2009novel}.

The novelty of this study presents itself when examining the impact and rehydration response using polarisation sensitive metrics. Following an EL1 impact, no significant change in birefringence was found between cartilage health states and the coefficient of restitution for EL1 impact events were similar in magnitude. Physiologically, the movement of water in the matrix is strain rate-dependent and hence articular cartilage is able to exploit its full viscoelastic properties to withstand the impact. As a result, there is little lasting disruption to the collagen network other than the formation of cracks at the high stress points (edges) of the impact cylinder. 

Extensive research into the pathogenesis of OA indicates that a reduction in proteoglycan content\cite{tiderius2003delayed}, swelling of the matrix\cite{mankin1975water,bollet1966biochemical}, and `de-structuring' of the collagen network\cite{thambyah2012macro} occurs in the very early stages of degeneration, well before clinical OA symptoms arise. We believe the combination of the above changes result in degenerate tissue no longer being able to fully utilise the fluid phase during high energy impacts and thus the intrafibrillar water is exuded from the matrix with little resistance. This is evidenced by the significant change in birefringence of the mildly degenerate tissue post impact and the return toward the pre-impact state following rehydration. Therefore, the already compromised collagen network bears the burden of the impact and drastically deforms to dissipate the impact energy and ultimately leads to a disturbed matrix exhibiting strong birefringent banding patterns present following EL2 impact events. These results support the notion that degenerate cartilage has an increased vulnerability to traumatic damage/injury\cite{workman2017influence}. The reduction in optical activity during the 3 hour rehydration period is particularly intriguing. From figure~\ref{fig:PSOCT}, it appears that rehydration of the degenerate matrix following an EL2 impact allows partial recovery of the collagen orientation. Unloaded, the collagen network returns toward the traditional Benninghoff arcade structure~\cite{benninghoff1925form} through pressurisation of the collagen network due to the Gibbs-Donnan osmotic swelling~\cite{maroudas1976balance}.  Such dynamic recovery inherently limits the diagnostic accuracy of birefringence as an absolute metric for assessing traumatic injury as there is little doubt the integrity of the tissue has been compromised during impact. However, it does demonstrate PS-OCT has great potential for providing insight into the complex mechanical response of articular to impact and extract quantitative OCT metrics that can predict the physiological state of cartilage in real-time.

Technical limitations included supervision and intervention during the semi-automated segmentation algorithm. Furthermore classification of a continuous disease into discrete categories coupled with macroscopic classification technique leads to some degree of variation, particularly in the degenerate (G2) sample groups. The larger variances present in the quantitative analysis of degenerate samples reflect this. On a biological note, bovine cartilage was chosen as it provides a near-limitless supply of articular cartilage exhibiting early stage degeneration and has been validated as an animal model for early OA~\cite{hargrave2013bovineModel}. The availability of the tissue makes it particularly suitable for exploratory studies such as this one. However, that in itself is one of the main limitations as bovine cartilage appears optically different to the gold standard: human cartilage. Therefore, the translatability of the results in the human cartilage domain remains in question. Future studies examining such species specific differences observed using PS-OCT would allow clinically-relevant large sample size bovine studies to take place to overcome the limitations often associated with human cartilage studies.

\section{Conclusion}
This present study investigated the impact response of healthy and degenerate bovine articular cartilage to moderate and high energy impacts. Combining the mechanical and imaging data, we conclude that PS-OCT is able to detect mechanically significant microstructural changes in mildly degenerate tissue. The real-time and non-invasive nature of PS-OCT thus presents itself as an ideal candidate for cartilage degeneration assessment and the point-and-shoot nature of the modality allows important physiological time-based changes to be examined.
 
\section{Disclosures}
The authors declare no conflicts of interest.
\section{Acknowledgements}
The authors would like to acknowledge funding from Marsden Fund and Royal Society of New Zealand (UoA1509) which made this research possible.


\bibliography{article}   
\bibliographystyle{spiejour}   

\end{spacing}
\end{document}